# Field emission in ultrathin PdSe$_2$ back-gated transistors

A. Di Bartolomeo,[1,2,a)] A. Pelella,[1,2] F. Urban,[1,2] A. Grillo,[1,2] L. Iemmo,[1,2] M. Passacantando,[3] X. Liu,[4] F. Giubileo [2]

[1]*Department of Physics, University of Salerno, via Giovanni Paolo II, Fisciano, 84084, Italy*

[2]*CNR-SPIN, via Giovanni Paolo II, Fisciano, 84084, Italy*

[3]*Department of Physical and Chemical Sciences, University of L'Aquila, and CNR-SPIN L'Aquila, via Vetoio, Coppito 67100, Italy*

[4]*National Laboratory of Solid State Microstructures, School of Physics, Collaborative Innovation Center of Advanced Microstructures, Nanjing University, Nanjing, 210093, China*

[a)]*Author to whom correspondence should be addressed. Electronic mail: adibartolomeo@unisa.it.*

We study the electrical transport in back-gate field-effect transistors with ultrathin palladium diselenide (PdSe$_2$) channel. The devices are normally-on and exhibit dominant n-type conduction at low pressure. The electron conduction, combined with the sharp edge and the workfunction decreasing with the number of layers, opens the way to applications of PdSe$_2$ nanosheets in vacuum electronics. In this work, we demonstrate field emission from few-layer PdSe$_2$ nanosheets with current up to the $\mu A$ and turn-on field below $100 \, V/\mu m$, thus extending the plethora of applications of this recently isolated pentagonal layered material.

The material research of the last decade has been largely dominated by two-dimensional (2D) transition metal dichalcogenides (TMDs), which have been explored for numerous applications in electronics, optoelectronics, sensors, catalysis, etc.[1–4] More recently, the TMDs based on noble metals of group 10, such as $PtS_2$, $PtSe_2$, $PdS_2$ and $PdSe_2$, have attracted considerable interest for the strong layer-dependent properties and interlayer interactions arising from the d-orbitals nearly fully occupied and the highly hybridized p$_z$ orbital of the interlayer chalcogen atoms[5,6]. Layer-dependent tunable bandgap, air stability, anisotropy and relatively high carrier mobility are winning properties of such materials[6].

---

[a)] Author to whom correspondence should be addressed. Electronic mail: adibartolomeo@unisa.it.

$PdSe_2$ has been the first isolated layered material with pentagonal structure that is stable in air[7,8]. The monolayer has a puckered morphology where each $Pd$ atom is coordinated with four $Se$ atoms and exhibits an indirect bandgap of $\sim 1.3\ eV$ which vanishes with the increasing number of layers up to the metallic behavior of the bulk material. Owing to the tunable and narrow bandgap, multi-layer $PdSe_2$ has been utilized for long-wave[9] or ultra-broadband[10] infrared photodetectors at room-temperature with highly sensitivity[11–14]. The visible-light optical absorption, combined with the carrier effective mass and mobility that change along different transport directions, indicate $PdSe_2$ as a good material for electron/hole separation, suitable for photovoltaic applications[15,16]. The carrier mobility in the order of $100\frac{cm^2}{Vs}$ or more has been demonstrated in field-effect transistors (FETs) with ambipolar characteristics and dominant polarity tunable by pressure or chemical doping[4,17]. The low lattice thermal conductivity at room temperature enables promising thermoelectric applications[18,19]. Due to the high bandgap variability and air stability, 2D $PdSe_2$ nanosheets have been also successfully used as a saturable absorber for generation of Q-switched laser pulses[20].

Motivated by the attractive properties and wide applications, we investigated 2D $PdSe_2$ nanosheets for field emission, i.e. for electron extraction under the application of a high electric field. Field emission (FE) offers significant scientific interests in material science and is utilized for electron microscopy, electron spectroscopy, e-beam lithography as well as in vacuum electronics for displays and microwave generation or for x-ray tubes[21–25]. The extraction of electrons is favored from low work function metals or semiconductors (the work function is related to the height of the energy barrier that the electrons have to tunnel through) and from sharply shaped surfaces due to their field-enhancing effect[26,27]. The expected $PdSe_2$ advantages, besides the air stability, are related to the sharp edges of the ultrathin flakes, the intrinsic n-type conduction caused by $Se$ vacancies[28,29] and to the work function that decreases with the reducing number of layers[11]. Furthermore, FE from few-layer $PdSe_2$ nanosheets is still an experimentally unexplored phenomenon.

In this work we report the measurement and the characterization of the field emission current from exfoliated $PdSe_2$ flakes. Besides its fundamental interest, our study suggests another promising application of 2D $PdSe_2$.

The $PdSe_2$ nanosheets were obtained from bulk $PdSe_2$ single-crystals using the standard mechanical exfoliation method by adhesive tape. For the synthesis of bulk $PdSe_2$, selenium powder (99.999 %) and palladium powder (99.95 %) were mixed in an atomic ratio of $Se:Pd = 2:1$, compressed into tablets, and placed into a quartz tube sealed under $10^{-5}\ mbar$. Collocated inside a muffled furnace, the quartz tube was gradually heated up and then kept to the synthesis temperature of 850 °C for 70 $h$. After natural cooling down of the furnace to room temperature, the obtained poly-crystalline samples of $PdSe_2$ were mixed with $Se$ powder in a mass ratio of $PdSe_2:Se = 1:4$ and sealed in another evacuated quartz tube to repeat the above high temperature annealing process.

The exfoliated nanosheets were transferred onto a substrate of degenerately doped p-type silicon covered with 300 $nm$ thick $SiO_2$. Optical microscopy was applied to select flakes with thickness of $15 - 20\ nm$ which were subsequently contacted with $5/40\ nm\ Pd/Au$ bilayers through electron-beam lithography, metal evaporation and lift-off. The chemical composition of the nanosheets investigated by energy dispersive X-ray spectroscopy revealed a $Pd:Se$ atomic ratio close to $1:2$, while X-ray diffractometry and Raman spectroscopy confirmed a layered crystal structure along the $c$-axis, as reported elsewhere[4] (the unit cell of $PdSe_2$ is orthorhombic with space group Pbca[7,29,30]).

The scanning-electron (SEM) and atomic-force (AFM) microscope images of a selected $PdSe_2$ nanosheet contacted by several metal leads, at a distance of about 950 $nm$ from each other, is shown in Figure 1. The AFM step height measurement (Figures 1(b) and 1(c)) indicates a ~17 $nm$ thick nanosheet, which corresponds to ~40 atomic layers (assuming the theoretical monolayer thickness of 0.41 $nm$[17]).

The electrical measurements were performed at room temperature and under controlled pressure inside a SEM chamber (ZEISS, LEO 1530). We contacted the samples using piezo-driven tungsten tips (W-tips) which are moved in steps of 5 $nm$ and positioned in the desired location with the help of the SEM imaging. The W-tips are connected to a Keithley 4200 semiconductor analyzer system. For the electrical measurements we used a three-terminal configuration, with different combinations of the metal leads as the source and drain and the Si substrate as the back gate of a field effect transistor (a schematic is shown in Figure 1(d)).

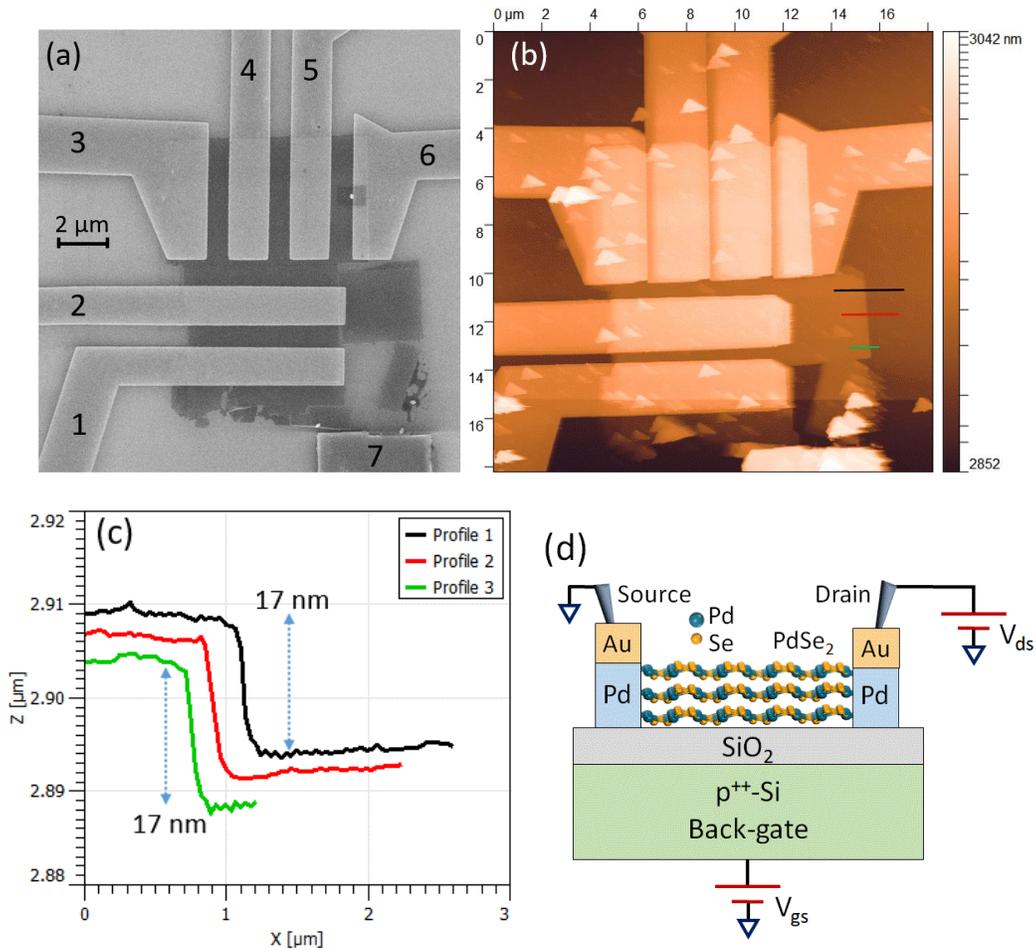

FIG. 1. (a) SEM and (b) AFM images of a $PdSe_2$ flake contacted with $Pd/Au$ leads. (c) Edge profile of the nanosheet showing a step height of $\sim 17\ nm$. (d) Schematic cross-section and biasing of the $PdSe_2$ field-effect device in common-source configuration.

The linear and symmetric output curves ($I_{ds} - V_{ds}$ drain current versus voltage for stepping gate voltage $V_{gs}$ in common source configuration) of Figure 2(a), referred to contacts labelled 4 and 5, indicate ohmic contacts. Figure 2(b) shows a typical set of transfer curves ($I_{ds} - V_{gs}$ drain current vs gate voltage at a given drain bias) for the gate voltage swept back and forth over increasing ranges. The measurements were performed after keeping the sample at low pressure ($< 10^{-6}\ Torr$) for several days to help removing adsorbates such as $O_2, H_2O$, etc. which could strongly affect the $PdSe_2$ conductivity[4,28]. The device is in the on-state at zero gate, i.e. behaves as a normally-on transistor, and has a dominant n–type behavior. The n-type conduction of $PdSe_2$ can be attributed to intrinsic defects, such as selenium vacancies[28], whose presence is confirmed by the clockwise[31] and widening hysteresis[32] with the increasing gate bias range. The transistor has two orders of magnitude on/off ratio, a negative threshold voltage, $V_{th} < -10\ V$, and a maximum field-effect electron mobility $\mu = \frac{L}{W} \frac{1}{C_{SiO_2}} \frac{1}{V_{ds}} \frac{dI_{ds}}{dV_{gs}} \sim 20\ cm^2 V^{-1} s^{-1}$ (here $L$ and $W$ are the channel length and width, respectively, and $C_{SiO_2} = 11\ nF\ cm^{-2}$ is the capacitance per unit area of the 300 $nm$ $SiO_2$ gate dielectric). The electron mobility in $PdSe_2$ has been shown to decrease with the increasing number of layers[7]; the achieved value is consistent with that reported for devices with similar thickness[7,17,33].

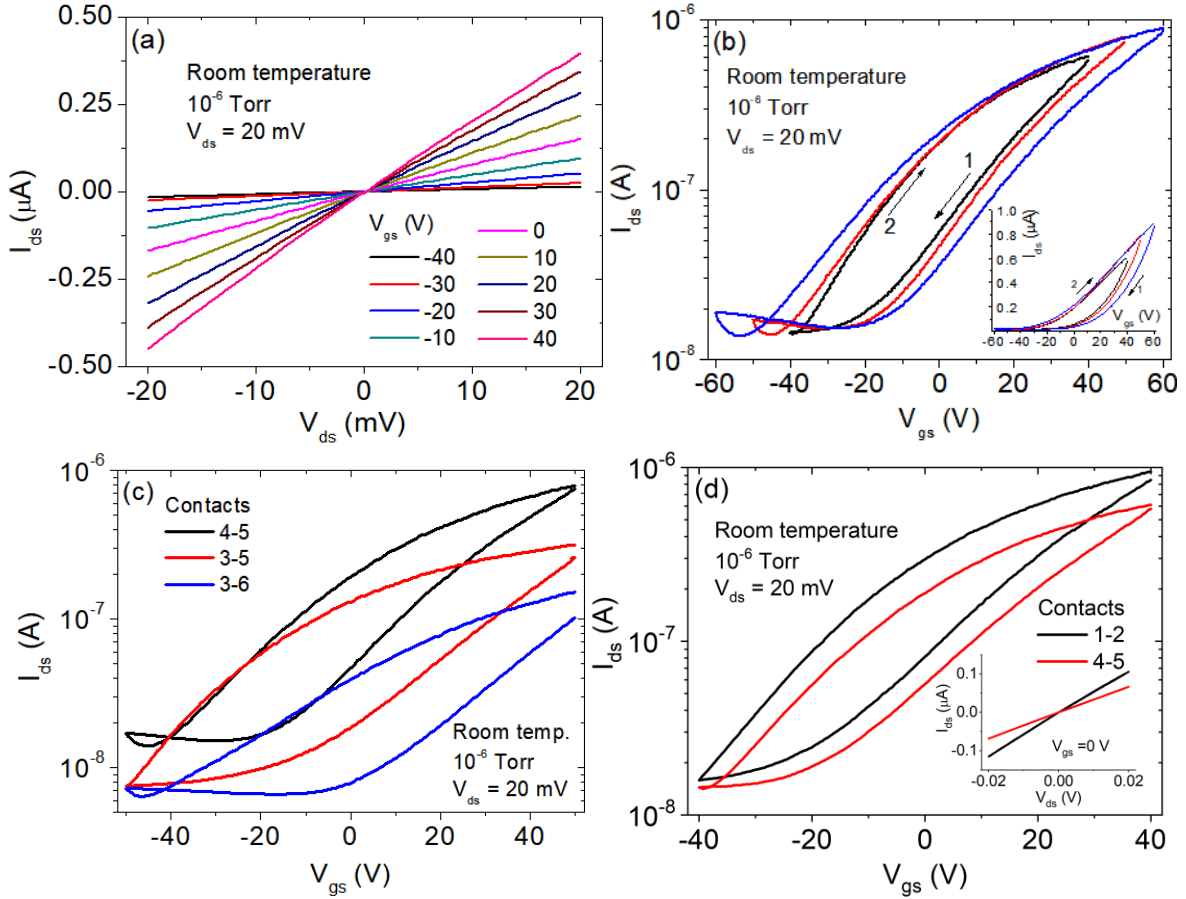

FIG. 2. (a) Output and (b) transfer characteristics measured between leads 4 and 5 (channel length $L = 0.96\ \mu m$ and width $W = 5.20\ \mu m$). The three transfer characteristics correspond to increasing gate voltage ranges. The bottom inset shows the transfer curves on linear scale. (c) Transfer curves between different contacts combinations (contacts 3 and 5: $L = 3.60\ \mu m$ and $W = 5.20\ \mu m$; contacts 3 and 6: $L = 6.60\ \mu m$ and $W = 5.20\ \mu m$). (d) Transfer (and output, as inset) characteristics between leads 1 and 2 (channel length $L = 0.94\ \mu m$ and width $W = 6.50\ \mu m$) and leads 4 and 5.

Remarkably, Figure 2(b) shows the appearance of hole conductance (ambipolar behavior) for $V_{gs} < -50\ V$. However, the limitation imposed to the maximum gate voltage to avoid $SiO_2$ breakdown, prevented further characterization of the p-type conduction. To assess the repeatability of the reported behavior we checked different contact combinations, finding similar results even when the channel included intermediate floating metal leads (Figure 2(c)). Similarly, the comparison of the transfer characteristics for contacts 1-2 and 4-5, that correspond to transport along two perpendicular crystallographic axes, do not evidence any appreciable difference as the higher current between contacts 1-2 can be easily attributed to the larger and shorter channel.

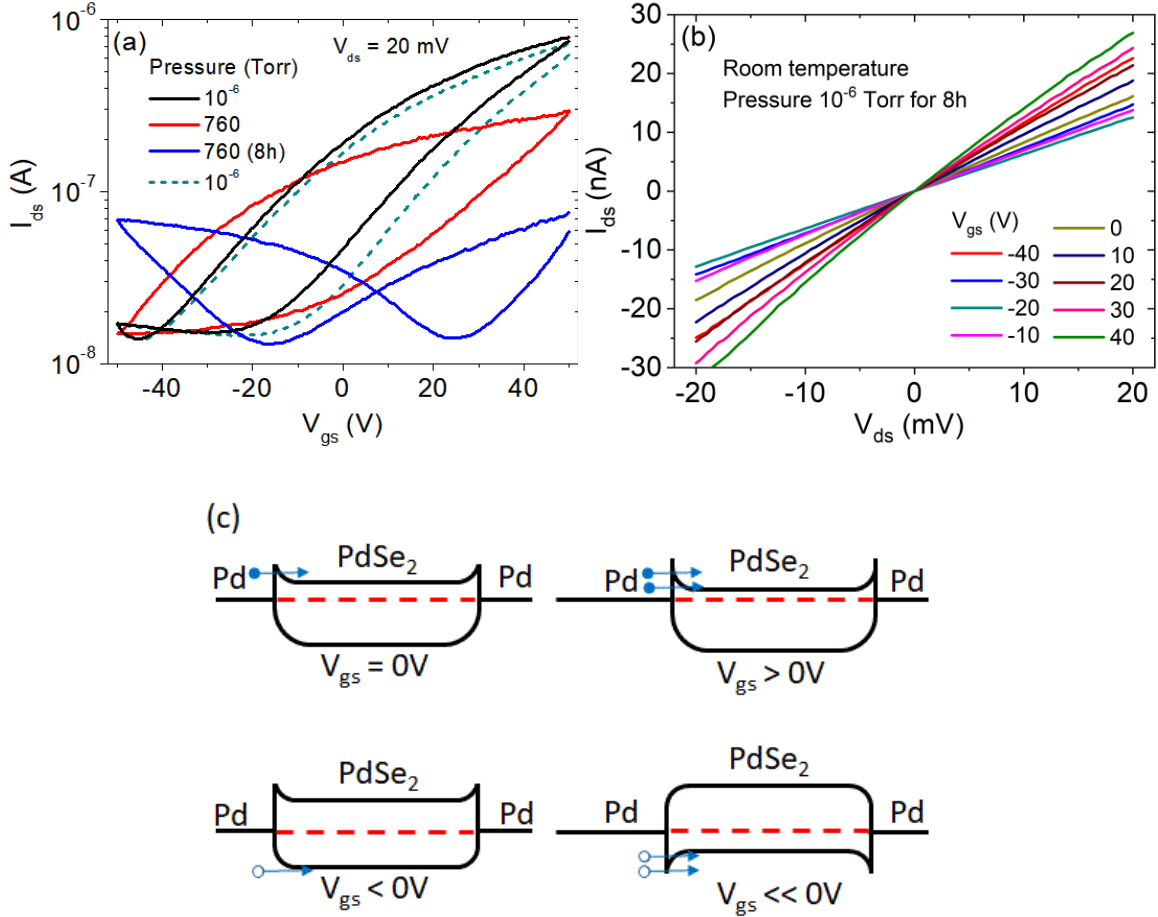

FIG. 3. (a) Transfer characteristics measured between leads 4 and 5 in vacuum, after exposure to air and after pump-down to pressure $< 10^{-6}\ Torr$. (b) Output characteristics in air after $8h$ exposure to air. (c) Energy band diagrams at different gate voltages.

We confirmed that the electrical behavior of $PdSe_2$ is strongly influenced by the environment[4,28], when at the end of our experiment we exposed the device to air. Figure 3(a) shows that the electron current gradually decreases and a clear ambipolar behavior appears after an overnight exposure to air, while the ohmic behavior of the contacts is maintained (Figure 3(b)). However, consistently with previous studies on the reversibility of short-term aging of $PdSe_2$ in air[4,28,34], Figure 3(a) shows that the air induced change is reversible: the device returns to its n-type dominant behavior after the re-establishment of the $10^{-6}\ Torr$ pressure.

The observed electrical behavior can easily be understood considering that the multilayer nanosheet has low bandgap around $0.1\ eV$ or less[7,8,33] and that $Pd$ has higher work function $(5.20 - 5.95\ eV)$[35,36] than $PdSe_2$ ($5.16 - 5.20\ eV$ for multilayers[11,12]). Consequently, as clarified by the schematic band diagrams of Figure 3(c), for intrinsic n-type doping, there is a band bending that favors electron conduction at zero and positive gate; the application of a negative gate voltage suppresses the n-type conduction and enables hole conduction at higher negative gate biases.

Considering the sharp edges and taking advantage of the workfunction that decreases with the reducing number of layers down to $\sim 4.30\ eV$ for the monolayer[11,12], the 2D form of $PdSe_2$ can be a good candidate for FE applications. To investigate FE from few-layer $PdSe_2$ nanosheets, we selected flakes protruding from the metal leads. Figure 4(a) shows an example of a flake emerging from lead 1 with thickness $\sim 4\ nm$ (about 10 layers) as measured by AFM. For the FE measurements, we positioned one of the W-tips on the metal lead (cathode) and approached the $PdSe_2$ flake with the second W-tip (anode). The current-voltage (I-V) characteristic obtained between lead 1 and the anode tip in electric contact with the nanosheet is shown in Figure 4(b). The slightly rectifying p-type behavior points to the formation of a low Schottky barrier between the W-tip and the $PdSe_2$ nanosheet[37]. Despite the rough contacting, the I-V curve appears smooth and a current up to the $\mu A$ can be attained at $V = 60\ V$.

Starting from the contact condition, we retracted the anode W-tip to given distances from the edge of the $PdSe_2$ nanosheet and we performed voltage sweeps up to $100\ V$, while monitoring the current. Figure 4(c) shows a typical sequence where from the contact condition (magenta curve) we retracted the anode tip to distances $\sim 100\ nm$ and $\sim 310\ nm$ (dark cyan and red curves, respectively). At $d \sim 100\ nm$, the I-V behavior mimics the one of the contact condition: Likely, a sort of loose electric contact has established between the flake and the tip as result of electrostatic attraction. Instead, at $d \sim 310\ nm$ an exponentially growing current, typical of the FE, emerges from the setup noise floor (obtained with the anode W-tip far away from the nanosheet) at $V \sim 45\ V$. This voltage corresponds to a turn-on electric field $E_{to} = \frac{V}{kd} \sim 90\ V/\mu m$ ($k = 1.6$ is the tip correction factor to account for the spherical geometry of the tip[38,39]). We repeated the same sequence, at a near location (that shown in Figure 4(a)), with the anode tip in contact and at $70\ nm, 200\ nm, 300\ nm$ distances, respectively, confirming the appearance of FE with a turn-on electric field increasing from ~ $60\,V/\mu m$ at $70\,nm$ to ~ $90\,V/\mu m$ at $350\,nm$. We did not observe any apparent modification of the $PdSe_2$ nanosheet after the FE measurements.

Although recently challenged[40], the FE current is usually analyzed using the simplified Fowler–Nordheim (FN) equation[41]:

$$I = S\,a\frac{E_S^2}{\Phi}\cdot exp\left(-b\frac{\Phi^{3/2}}{E_S}\right), \qquad (1)$$

where $\Phi(eV)$ is the work function of the emitting material, $S(cm^2)$ is the emitting area, $E_S\,(V/cm)$ is the local electric field, and $a$ ($1.54\times 10^{-6}\,AV^{-2}eV$) and $b$ ($6.83\times 10^7\,Vcm^{-1}eV^{-3/2}$) are dimensional constants. The local electric field is $E_S = \beta\,V/d$, where V is the applied potential, $d$ is the cathode–anode distance, and $\beta$ is the so-called field enhancement factor. Such a factor is related to the geometry of the cathode, with the sharper surfaces yielding the higher $\beta$, and can attain values of several thousands[24,42,43]. According to the FN theory, the slope of the straight-line fitting the $ln(I/V^2)$ vs. $1/V$ FN plot can be used to estimate $\beta$.

It has been recently suggested that when the material is very thin, but not necessarily two-dimensional, due to discrete bound states, many basic assumptions of FN model become invalid and a modified FE scaling law should be used[44]:

$$I \propto exp\left(-b\frac{\Phi^{3/2}}{E_S}\right), \qquad (2)$$

Figures 4(e) and 4(f) show eq. (1) and eq. (2) fittings. The good agreement of both models with the experimental data at $d_1 \sim 200\,nm$ and $d_2 \sim 350\,nm$ provides evidence of the establishment of a FE regime but does not allow a clear discrimination between the two models. The magenta curve at a distance of $70\,nm$ deviates from the linear behavior and is something in between FE and physical electric contact (dark cyan curve).

Using the data of Figure 4(e), we can estimate a field enhancement factor of 5 at $200\,nm$ and 40 at $350\,nm$, assuming $\Phi = 4.70\,eV$ as the work function of the $4\,nm$ flake[12]. The increasing turn-on field and field enhancement factor with the anode-cathode distance is typical of nanoshaped emitters and has been discussed elsewhere[45–49]. We note also that the field emission figures-of-merit, namely turn-on field and field enhancement factor, are comparable to those obtained for graphene[50–52] or other transition metal dichalcogenides, such as $MoS_2$[53,54] or $WSe_2$[55], under similar experimental conditions.

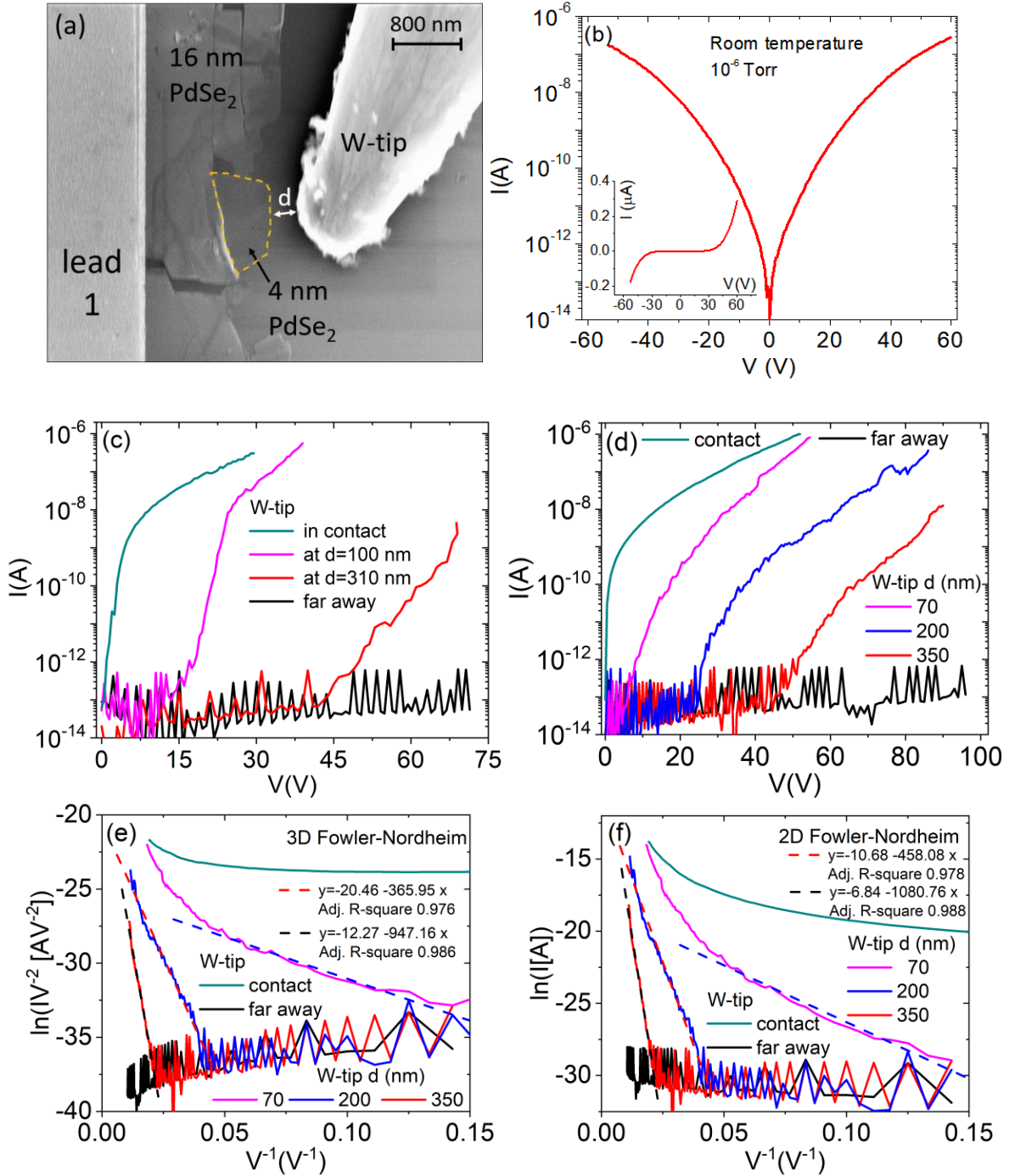

FIG. 4. (a) SEM image of a $PdSe_2$ nanosheet (cathode) protruding from metal lead 1 and the anode W-tip used for field emission measurements. (b) I-V characteristic with the anode W-tip in physical contact with the $PdSe_2$ nanosheet. The inset shows the I-V curve on linear scale. (c) and (d) I-V curves with the anode W-tip at growing distances from the $PdSe_2$ nanosheet showing the evolution from electric contact to FE regime. The substrate back-gate was grounded during all the above measurements. (e) and (f) Fowler-Nordheim plots obtained from eq. (1) and eq. (2) models.

In summary we have fabricated $PdSe_2$ nanosheets that we have electrically characterized using back-gated field effect transistors. Taking advantage of the intrinsic n-type conductivity, enhanced by the low pressure, of the sharp edges of flakes and of the layer-controllable work function, we have demonstrated that a high FE current up to the $\mu A$ can be extracted from 2D $PdSe_2$ nanosheets. Our study makes a step ahead towards the understanding and the application of $PdSe_2$ in its 2D form.